\documentclass{iaus}
\usepackage{graphicx}

\newcommand{\gta}{\gtrsim}

\newcommand{\kms}{\>{\rm km}\,{\rm s}^{-1}}
\newcommand{\gyr}{\>{\rm Gyr}}
\newcommand{\myr}{\>{\rm Myr}}

\newcommand{\pc}{\>{\rm pc}}

\newcommand{\msun}{\>{\rm M_{\odot}}}

\newcommand{\bdm}{\begin{displaymath}}
\newcommand{\edm}{\end{displaymath}}
\newcommand{\beq}{\begin{equation}}
\newcommand{\eeq}{\end{equation}}
\newcommand{\bit}{\begin{itemize}}
\newcommand{\eit}{\end{itemize}}
\newcommand{\ben}{\begin{enumerate}}
\newcommand{\een}{\end{enumerate}}
\newcommand{\bfi}{\begin{figure}[htb]}
\newcommand{\bpfi}{\begin{figure}[p]}

\title[Nuclear Star Clusters] %% give here short title %%
{Nuclear Star Clusters}

\author[T. B\"oker]   %% give here short author list %%
{Torsten B\"oker}
\affiliation{European Space Agency, Keplerlaan 1, 
200AG Noordwijk, the Netherlands \\ email: {\tt tboeker@rssd.esa.int} }

\pubyear{2009}
\volume{266}  %% insert here IAU Symposium No.
\pagerange{119--126}
\jname{Star Clusters: Galactic Building Blocks Through Time And Space.}
\editors{Richard de Grijs \& Jacques Lepine, eds.}
\begin{document}

\maketitle

\begin{abstract}
The centers of most galaxies in the local universe are occupied by compact, barely resolved 
sources. Based on their structural properties, position in the fundamental plane, and integrated spectra, these sources 
clearly have a stellar origin. They are therefore called "nuclear star clusters" (NCs) or "stellar 
nuclei". NCs are found in galaxies of all Hubble types, suggesting that their formation is intricately
linked to galaxy evolution. Here, I review some recent studies of NCs, describe ideas for their
formation and subsequent growth, and touch on their possible evolutionary 
connection with both supermassive black holes and globular clusters.
\end{abstract}

\firstsection % if your document starts with a section,
\section{Introduction}
The nuclei of galaxies are bound to provide ``special'' physical conditions because they 
are located at the bottom of the potential well of their host galaxies. This unique
location manifests itself in various distinctive phenomena such as super-massive
black holes (SMBHs), active galactic nuclei (AGN), central starbursts, or extreme 
stellar densities. The evolution 
of galactic nuclei is closely linked to that of their host galaxies, as inferred from a 
number of global-to-nucleus scaling relations discovered in the last decade.

Recently, observational and theoretical interest has been refocused
onto the compact and massive star clusters found in the nuclei of galaxies of all Hubble 
types. These ``nuclear star clusters'' (NCs) are intriguing
objects that are linked to a number of research areas: i) they are a promising
environment for the formation of massive black holes because of their extreme stellar 
density, ii) they may also constitute the progenitors of at least some 
halo globular clusters via ``NC capture'' following the tidal disruption 
of a satellite galaxy, and iii) their formation process is influenced by (and 
important for) the central potential, which in turn governs the
secular evolution of their host galaxies.

In what follows, I briefly summarize what has been 
learned about NCs over the last few years, describe some proposed 
formation mechanisms of NCs, and discuss the new paradigm of ``central massive
objects'' which links NCs with SMBHs in galactic nuclei. Lastly, I briefly mention
a scenario in which NCs may be the progenitors of (some) globular clusters.

\section{Properties of Nuclear Star Clusters} \label{sec:properties}
Extragalactic star clusters are compact sources, and in general, their study 
requires high spatial resolution afforded only by the {\it Hubble Space Telescope}
or large ground-based telescopes using adaptive optics. 
Over the last decade, a number of studies - both via imaging and
spectroscopic observations - have contributed to the following picture of NCs:

\vspace*{2mm}
\noindent {\bf 1)} NCs are common: the fraction of galaxies with an unambiguous NC detection
is 75\% in late-type (Scd-Sm) spirals (\cite{boe02}), 50\% in earlier-type (Sa-Sc)
spirals (\cite{car97}), and 70\% in spheroidal (E \& S0) galaxies (\cite{cot06}). All these
numbers are likely lower limits, although for different reasons. In the latest-type
disks, it is sometimes not trivial to locate the galaxy center unambiguously so that
no particular source can be identified with it. In contrast, many early-type galaxies
have very steep surface brightness profiles (SBPs) that make it difficult to detect
even luminous clusters against this bright background.

\vspace*{2mm}
\noindent {\bf 2)} NCs are much more luminous than ``normal'' globular clusters (GCs). 
With typical absolute I-band magnitudes between -14 and -10 (\cite{boe02,cot06}), they 
are roughly 40 times more luminous than the average MW globular cluster (\cite{har96}).

\vspace*{2mm}
\noindent {\bf 3)} However, NCs are as compact as MW GCs. Their half-light radius 
typically is $\rm 2-5\,pc$, independent of galaxy type (\cite{boe04,geh02,cot06}). 

\vspace*{2mm}
\noindent {\bf 4)} Despite their compactness, NCs are very massive: their typical 
dynamical mass is $10^6 - 10^7\msun$ (\cite{wal05}) i.e. at the extreme high 
end of the GC mass function. 

\begin{figure}[t]
\centering
\includegraphics[width=7cm]{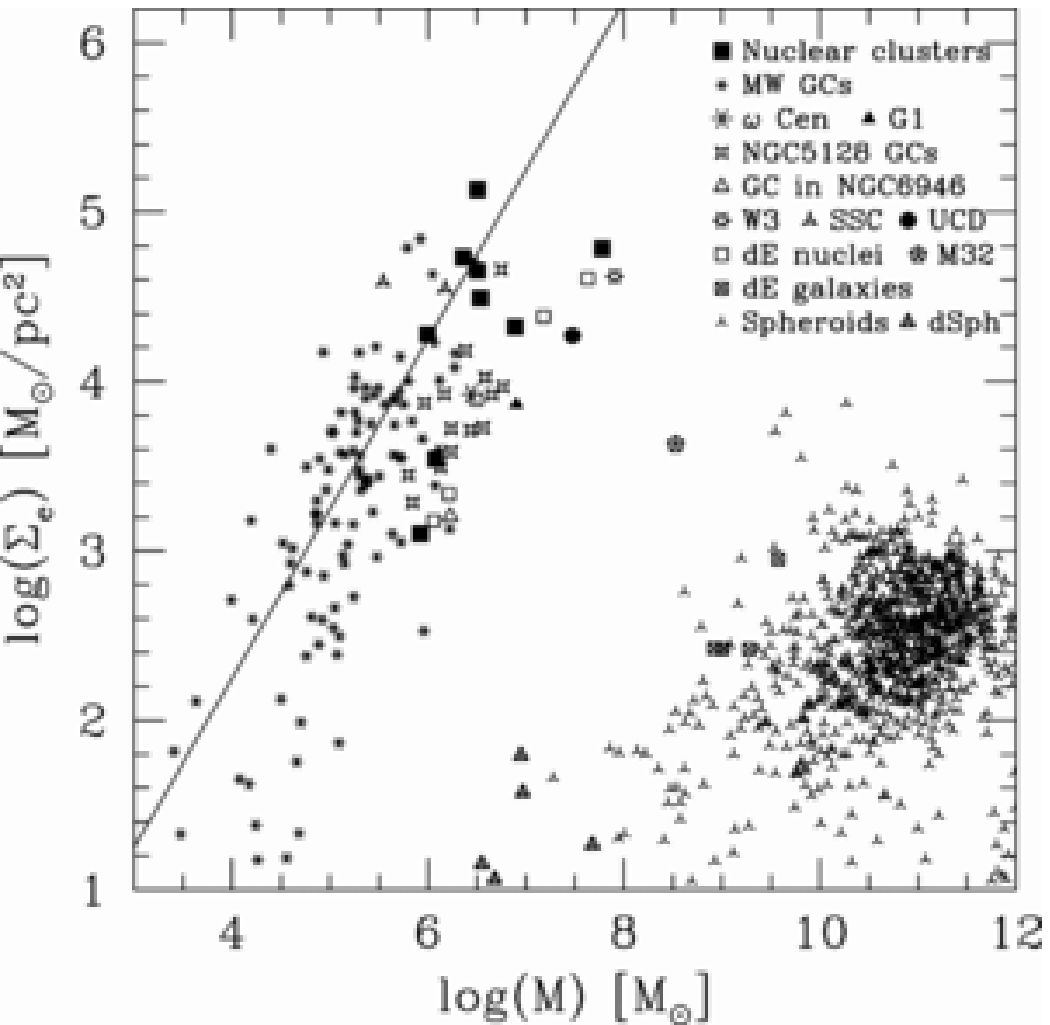}
\caption{Mean projected mass density of various stellar systems inside their 
effective radius $r_e$, plotted against their total mass.
This is similar to a face-on view of the fundamental plane. 
NSCs occupy the high end a region populated by other types of massive stellar clusters,
and are well separated from elliptical galaxies and spiral bulges. The solid line 
represents a constant cluster size, i.e. $r_e=3\pc$ (from \cite{wal05}).}
\label{fig:fp}
\end{figure}

\vspace*{2mm}
\noindent {\bf 5)} Their mass density clearly separates NCs from compact galaxy bulges. This
is demonstrated in Figure~\ref{fig:fp} which compares the mass and mass density of
NCs to that of other spheroidal stellar systems. The clear gap between bulges/ellipticals
on the one hand, and NCs on the other hand makes a direct evolutionary connection
between the two classes of objects unlikely.

\vspace*{2mm}
\noindent {\bf 6)} The star formation history of NCs is complex, as evidenced by the fact
that most NCs have stellar populations comprised of multiple generations of stars
(\cite{wal06,ros06}). While all NCs show evidence for an underlying old ($\gta 1\gyr$)
population of stars, most also have a young generation with ages below $100\myr$. This
is strong evidence that NCs experience frequent and repetitive star formation episodes
(\cite{wal06}). 

\vspace*{2mm}
\noindent {\bf 7)} NCs obey similar scaling relations with host galaxy properties
as do SMBHs. This finding has triggered a very active research area, but its implications 
are still to be understood fully, as discussed in more detail in  \S\,\ref{sec:cmo}.

\section{Possible Formation Mechanisms} \label{sec:formation}
There are a number of suggested formation scenarios for NSCs, and so far,
few have been ruled out. In principle, one can distinguish between two main
categories: a) migratory formation scenarios in which dense clusters form 
elsewhere in  the galaxy, and then fall into the center via dynamical friction or
other mechanisms (\cite{cap08,and08}), and b) in-situ cluster build-up via 
(possibly episodic) gas infall and  subsequent star formation within a few parsec 
from the galaxy center.

The processes that funnel gas onto NCs in nearby galaxies have recently been
studied in some detail, enabled by significant improvements to the sensitivity
and spatial resolution of mm-interferometers (e.g. \cite{sch06}, 2007). In general,
bar-shaped asymmetries in the disk potential can lead to prolonged influx of molecular
gas into the central few pc, thus providing the reservoir for an intense burst of
star formation, and leading to the rejuvenation of an existing NC.
The starburst is, however, self-regulating in the sense that mechanical
feedback from stellar winds and/or supernova explosions can expel the remaining
gas, and even temporarily change the gas flow pattern (\cite{sch08}). This scenario
naturally leads to episodic star formation, thus explaining the presence of
multiple stellar populations in NCs.

Less clear, however, are the reasons for why gas accumulates in the nucleus of a shallow
disk galaxy {\it in the absence} of a prominent central mass concentration, i.e.
how the ``seed clusters'' form initially. A few studies have attempted to provide 
an explanation for this puzzle. For example, \cite{mil04} suggests the magneto-rotational 
instability in a differentially rotating gas disk as a viable means to transport gas 
towards the nucleus and to support (semi)continuous star formation there.

More recently, \cite{ems08} have pointed out that the tidal field becomes compressive
in shallow density profiles, causing gas to collapse onto the nucleus of a disk galaxy. 
If correct, then NC formation is indeed expected to be a natural consequence of galaxy 
formation, which would go a long way towards explaining at least some of the observed
scaling relations between NCs and their host galaxies.

The question of when a particular NC (i.e. its ``seed'' cluster) has formed is 
equivalent to asking how old its oldest stars are. This question is extremely 
difficult to answer in all galaxy types, albeit for different reasons. In late-type 
spirals, for example, the NC nearly always contains a young stellar population
which dominates the spectrum and thus makes the detection of an underlying
older population challenging, not to mention its accurate age determination.

Early-type galaxies, on the other hand, have much steeper surface brightness profiles, 
and therefore a low contrast between NC and galaxy body. This makes spectroscopic 
studies of NCs in E's and S0's exceedingly difficult.
The few published studies have focused on the NCs of dE,N galaxies, and have
shown that even these can have stellar populations that are significantly younger
than the rest of the host galaxy (\cite{but05,chi07,kol09}).
In general, stellar population fits as well as the rather high dynamical 
mass-to-light ratios of NCs indicate that they contain a significant population 
of evolved (at least 1\,Gyr old) stars, i.e. they have been in place for a long time (\cite{wal05}).

\begin{figure}[t]
\begin{center}
\vspace*{-4.6cm}
\includegraphics[width=5.5in] {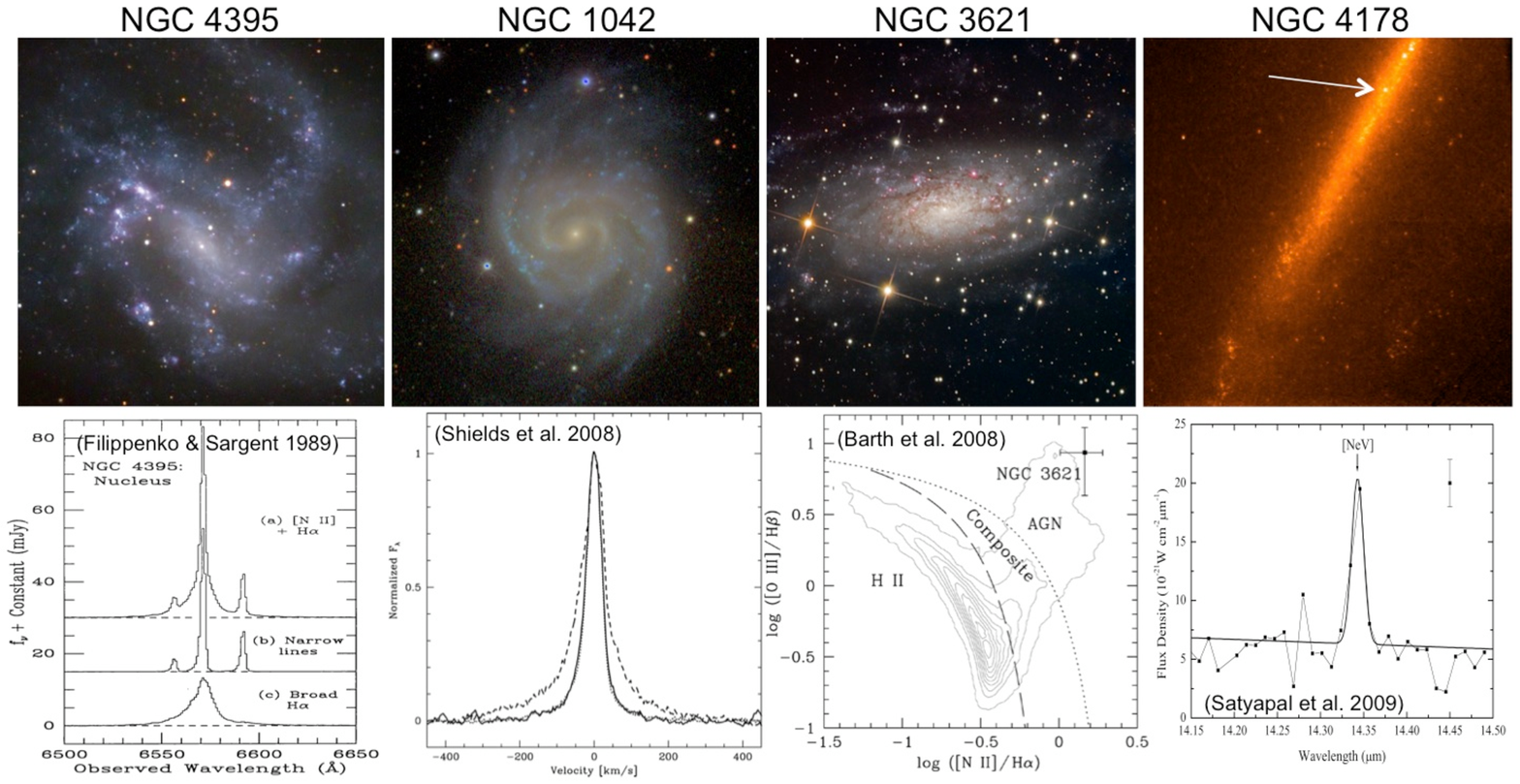} 
\vspace*{-6.6cm}
 \caption{Four bulgeless disk galaxies with evidence for an AGN
 (references in the lower panels). The case of NGC\,4395 has 
 long been thought to be unique, but detailed observations
 of other late-type disks have shown that such low-luminosity AGN are
 easily missed in optical surveys. Nevertheless, the AGN fraction in bulgeless 
 disks appears to be lower than in galaxies with more massive bulges.}
   \label{fig:AGN}
\end{center}
\end{figure}

\section{Central Black Holes and Nuclear Star Clusters} \label{sec:cmo}
A number of recent studies (\cite{ros06,weh06,fer06,bal07,gra07}) have demonstrated that NCs 
follow similar scaling relations with their host galaxies as do SMBHs, and typically 
extend these relations to lower SMBH masses. This has triggered speculation about
a common formation mechanism of NCs and SMBHs, which is governed mostly by
the mass of the host galaxy spheroid. The idea put forward is that NCs and SMBHs are
two incarnations of a ``central massive object'' (CMO). Galaxies above a certain mass 
threshold ($\approx 10^{10}\msun$) form predominantly SMBHs while lower mass 
galaxies form NCs.

On the other hand, it is well established that many galaxies contain {\it both} a NC and 
a SMBH (\cite{set08}). This is true even at the extreme late end of the Hubble 
sequence, i.e. in galaxies that have no bulge component at all. A famous example 
known for a long time is the ``mini-Seyfert'' NGC\,4395 (\cite{fil89}), but a number of
similar cases have been found recently, as demonstrated in Figure~\ref{fig:AGN}.
These AGN are often missed in in spectra taken with relatively wide apertures 
because the AGN signatures are faint compared to those of their surroundings,
especially in the presence of (circum)nuclear star formation. 

However, this does not imply that {\it all} NCs harbor a SMBH. On the contrary,
a recent survey of high-ionization [NeV] emission in late type galaxies (\cite{sat09}) 
indicates that AGN in bulgeless disks are indeed rare: only about 5\% of spirals
with Hubble type Sd-Sm objects show [NeV] emission. Interestingly, all these 
low-luminosity "mini-AGN" are found in galaxies that also host an NC, possibly
suggesting that the presence of a NC is necessary (but not sufficient) for 
the formation of a SMBH.

\begin{figure}[t]
\begin{center}
 \includegraphics[width=5.in]{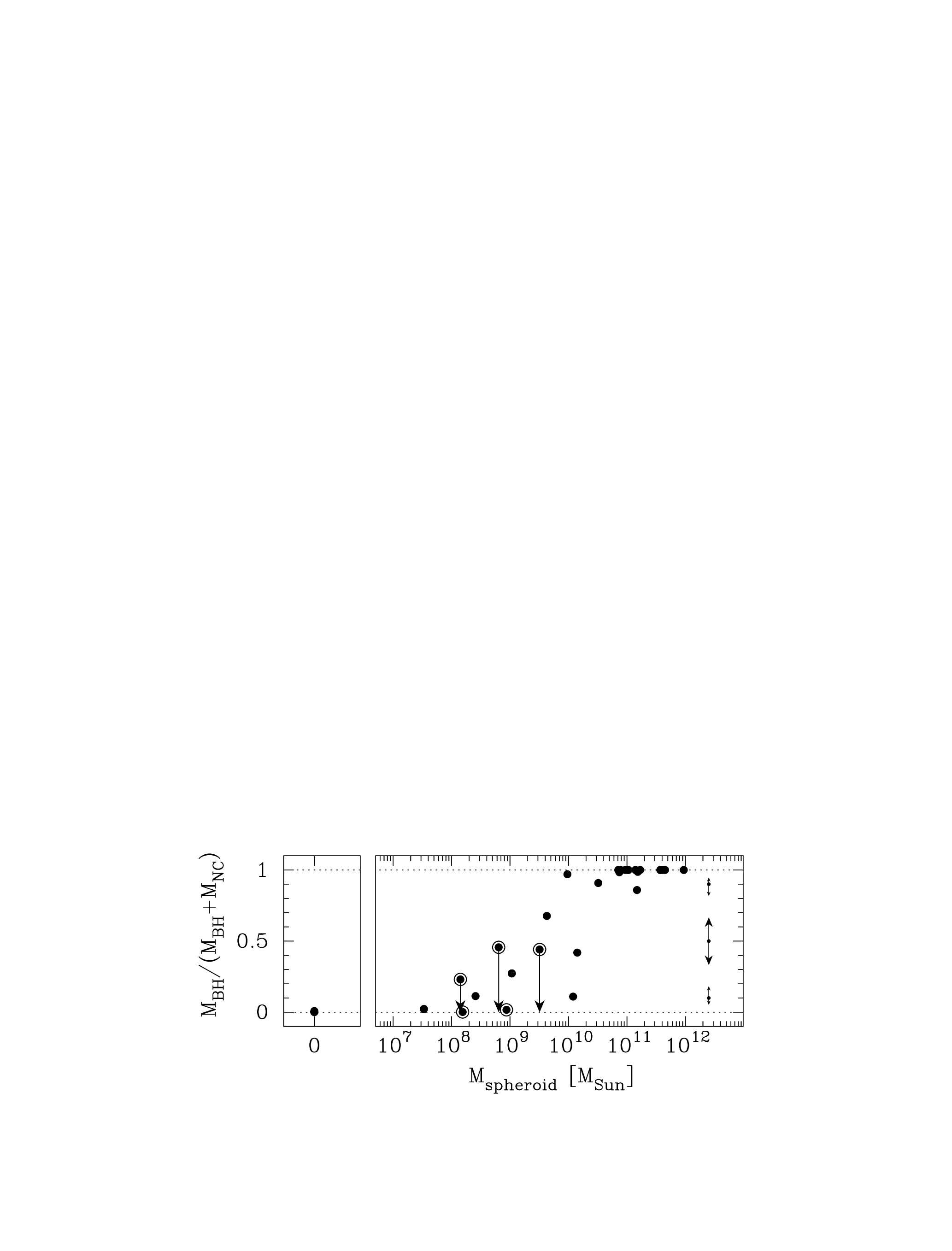} 
 \caption{
 The increasing dominance of the central BH over the NC of stars, 
 traced by the mass ratio $\rm M_{BH}/(M_{BH}+M_{NC})$ appears to 
 depend on the bulge mass $\rm M_{sph}$ of the host galaxy. The 
 leftmost data point indicates globular clusters which have zero bulge 
 mass. In contrast, the highest mass spheroids with the most massive
 BHs do not contain an NC (from \cite{gra09}). 
 }
   \label{fig:ratio}
\end{center}
\end{figure}

Why, then, do some NCs contain a SMBH, but not all? What is the mass
ratio between NC and SMBH, and how is it regulated? Important observational
constraints come from \cite{gra09} who have identified all galaxies with reliable
measurements of both NC mass and SMBH mass. They conclude that
the ratio $\rm M_{BH}/(M_{BH}+M_{NC})$ is a function of bulge mass, as
illustrated in Figure~\ref{fig:ratio}. Massive 
bulges only host SMBHs, but no NCs. At the other end of the spectrum, in "pure"
disk galaxies, the mass of the SMBH (if it exists at all) is negligible with respect 
to the NC mass. In between, there is a transition region
in which galaxies host both NCs and SMBHs with comparable masses.

A theoretical explanation for why this may be so has been offered by \cite{nay09}.
They speculate that ``competitive feedback'' between the SMBH and the gas inflow
feeding star formation during the NC buildup determines which of the two components
can grow more efficiently. The outcome of this ``race'' between the SMBH (which grows 
on the Salpeter timescale) and the NC (which should grow on the dynamical timescale 
regulating gas inflow) is decided by the bulge mass, i.e. the stellar velocity dispersion
$\sigma$. Below a value of $\sigma \approx 150\kms$, the BH cannot grow efficiently, 
above this value, it grows fast enough that its radiative feedback hinders NC growth. The
presence of a SMBH also has important consequences for the dynamical evolution
of a NC, since it prevents core collapse and might even disrupt the NC (\cite{mer09}).

One important question in this context which has been neglected so far, is why some 
galaxies apparently contain {\it neither} NC nor SMBH. In other words, how can a
galaxy avoid having a CMO? Progress along these lines will require a better understanding 
of the formation and survival of ``pure'' disk galaxies, a problem that is
still challenging for current models of structure formation.

\section{Nuclear Clusters as Precursors of Globular Clusters}
As mentioned in \S\,\ref{sec:formation}, most NCs have likely formed a long time ago.
In fact, some theories of structure formation suggest that already the first proto-galaxies
undergo rapid nucleation (\cite{cen01}), and form a dense star cluster in their center.
If these proto-galaxies are gas-rich, the NC will most likely experience multiple bursts of 
star formation similar to the present-day NCs in late-type disks. This process continues until
the proto-galaxy is destroyed in a merger. Because of its compactness and high stellar density,
the NC will survive the merger, and from that moment on will passively age in the halo of
the merger product. 

That this process does indeed occur is best demonstrated by the case of M\,54. This 
Milky Way GC is believed to be the nucleus of the Sagittarius dwarf galaxy (\cite{lay00}) 
which is currently being ``swallowed'' by the Milky Way. Another plausible example is
$\Omega$~Cen which has long been thought to be the remnant nucleus of an accreted 
dwarf galaxy because of it extreme mass and multiple stellar populations. 

As discussed in \cite{boe08}, assuming that GCs spent some part of their history
at the bottom of a galaxy's potential well might naturally explain the multiple 
stellar populations observed in a number of GCs in the Galaxy.
It is also consistent with the roughly constant specific frequency of GCs and
the observed universal mass fraction of GC systems in the local universe (\cite{mcl99}).

\end{document}